\documentclass[twocolumn,modern]{aastex631}
\usepackage{tabularx}

\usepackage{float}
\begin{document}

\title{A Measurement of the Water Abundance in the Atmosphere of the Hot Jupiter WASP-43b with High-resolution Cross-correlation Spectroscopy}

\author{Dare Bartelt}
\affiliation{Steward Observatory, University of Arizona, 933 N Cherry Ave, Tucson, AZ, USA}

\author{Megan Weiner Mansfield}
\affiliation{School of Earth and Space Exploration, Arizona State University, Tempe, AZ, USA}

\author{Michael R. Line}
\affiliation{School of Earth and Space Exploration, Arizona State University, Tempe, AZ, USA}

\author{Vivien Parmentier}
\affiliation{Universit\'e C\^ote d'Azur, Observatoire de la C\^ote d'Azur, CNRS, Laboratoire Lagrange, France}

\author{Luis Welbanks}
\affiliation{School of Earth and Space Exploration, Arizona State University, Tempe, AZ, USA}
\affiliation{NHFP Sagan Fellow}

\author{Elspeth K. H. Lee}
\affiliation{Center for Space and Habitability, University of Bern, Bern, Switzerland}

\author{Jorge Sanchez}
\affiliation{School of Earth and Space Exploration, Arizona State University, Tempe, AZ, USA}

\author{Arjun B. Savel}
\affiliation{Department of Astronomy, University of Maryland, College Park, MD, USA}

\author{Peter C. B. Smith}
\affiliation{School of Earth and Space Exploration, Arizona State University, Tempe, AZ, USA}

\author{Emily Rauscher}
\affiliation{Department of Astronomy, University of Michigan, Ann Arbor, MI, USA}

\author{Joost P. Wardenier}
\affiliation{Institut Trottier de Recherche sur les Exoplan\`etes, Universit\'e de Montr\'eal, Montr\'eal, Qu\'ebec, Canada}

\begin{abstract}
Measuring the abundances of carbon- and oxygen-bearing molecules has been a primary focus in studying the atmospheres of hot Jupiters, as doing so can help constrain the carbon-to-oxygen (C/O) ratio. The C/O ratio can help reveal the evolution and formation pathways of hot Jupiters and provide a strong understanding of the atmospheric composition. In the last decade, high-resolution spectral analyses have become increasingly useful in measuring precise abundances of several carbon- and oxygen-bearing molecules. This allows for a more precise constraint of the C/O ratio. We present four transits of the hot Jupiter WASP-43b observed between 1.45 $-$ 2.45 $\mu$m with the high-resolution Immersion GRating InfraRed Spectrometer (IGRINS) on the Gemini-S telescope. We detected H$_2$O at a signal-to-noise ratio (SNR) of 3.51. We tested for the presence of CH$_4$, CO, and CO$_2$, but we did not detect these carbon-bearing species. We ran a retrieval for all four molecules and obtained a water abundance of $\log_{10}(\text{H}_2\text{O}) = -2.24^{+0.57}_{-0.48}$. We obtained an upper limit on the C/O ratio of C/O $<$ 0.95. These findings are consistent with previous observations from the \textit{Hubble Space Telescope} and the \textit{James Webb Space Telescope}.
\end{abstract}


\section{Introduction} \label{sec:intro}
 The carbon-to-oxygen (C$/$O) ratio can reveal information about the formation and subsequent evolution of a planet \citep{2011ApJ...743L..16O, 2016ApJ...832...41M}. It can aid in determining where in the protoplanetary disk the planet formed and infer the disk chemistry \citep{Molliere2022}. It heavily influences the relative concentrations of spectroscopically dominant species, including water, methane, and hydrocarbons \citep{Madhusudhan_2012}. The C$/$O ratio can also hint at the total heavy element content of a giant planet's atmosphere and the influence of drifting and evaporating pebbles in the formation and evolution of these planets. The migration of a giant planet also has the potential to be inferred \citep{Schneider2021.1, Schneider2021.2}. It is important to note that the C/O ratio is just one piece of the puzzle in constraining the formation pathway of a planet. Determining abundances of refractory elements and volatile-to-refractory ratios (e.g., Si/H, O/Si, and C/Si) can provide important clues in unraveling planetary formation history \citep{Chachan_2023}. Accurate observational constraints on the C/O ratio ideally require measurements of all the primary carbon- and oxygen-bearing molecules, including H$_2$O, CO, CO$_2$, and CH$_4$. \\
 WASP-43b \citep{2011Hellier} is a 1.8 M$_{Jup}$ hot Jupiter that transits the K7V star, WASP-43, at an orbital distance of 0.015 AU with an orbital period of 0.81 days \citep{2011Hellier}. The close orbit suggests that WASP-43b is tidally locked to its host star \citep{2017Stevenson}. WASP-43b has been the subject of many atmospheric analyses, primarily using observations from the \textit{Hubble Space Telescope} \citep[\textit{HST},][]{Kreidberg_2014} and \textit{Spitzer} \citep{2014Blecic}. \cite{Kreidberg_2014} detected the presence of water in the transmission and emission spectra from the Wide Field Camera 3 (WFC3) on \textit{HST}. \textit{Spitzer} phase curves observed at 3.6 and 4.5 $\mu$m also detected water \citep{2017Stevenson}. These results are further supported by other independent analyses of these observations \citep{2019Morello, 2020MNRAS.493..106I, Tsiaras2018, Stevenson2017}. Although studies of the \textit{Spitzer} data have suggested the presence of CO, CO$_2$, and CH$_4$, the narrow wavelength range and the limited spectral resolving power have made it difficult to clearly identify these carbon-bearing molecules. More recent observations of WASP-43b by the \textit{James Webb Space Telescope} \citep[\textit{JWST},][]{Bell2024, Yang2024} showed signs of water vapor on both the dayside and nightside of the planet.\\ 
High-resolution cross-correlation spectroscopy (HRCCS) has emerged as a technique to obtain precise abundances of these key molecules and to better constrain the C$/$O ratio \citep{2010Natur.465.1049S,birkby2018exoplanetatmosphereshighspectral, Brogi2019, Line2021}. High-resolution spectroscopy resolves individual molecular lines, making it sensitive to fine features in atmospheric spectra. This technique has been used to identify volatile species in the atmospheres of several hot Jupiters, including WASP-76b \citep{deibert2023exogems, Tabernero_2021, WM2024}, WASP-77Ab \citep{Line2021}, WASP-18b \citep{Brogi2023}, WASP-121b \citep{wardenier2024phaseresolvingabsorptionsignatureswater} and HD 209458b \citep{Gandhi2019}. The first set of high-resolution dayside observations of WASP-43b was conducted by \cite{2023Lesjak} with the CRIRES$^+$ spectrograph. These observations detected H$_2$O and provided the first direct detection of CO in the atmosphere of this planet. These observations retrieved a super-solar C$/$O ratio of $0.78\pm0.09$. \\
In this paper, we measure the abundance of H$_2$O in the atmosphere of WASP-43b with four transits observed with the Immersion GRating INfrared Spectrometer (IGRINS) on the Gemini-S telescope. The high resolution of IGRINS (R $\sim$ 45,000) and its ability to observe in the 1.45 $-$ 2.45 $\mu$m wavelength range means it is sensitive to key carbon- and oxygen-bearing molecules, such as H$_2$O, CO, CO$_2$, and CH$_4$. In Section 2, we describe the observations and data reduction. In Section 3, we use the cross-correlation technique to infer the presence of H$_2$O and present non-detections of CO, CO$_2$, and CH$_4$ in the atmosphere. In Section 4, we use atmospheric retrievals to constrain the abundances of the detected gases and determine an upper limit for the C/O ratio. In Section 5, we discuss these observations in the context of previous studies of the atmospheric composition of WASP-43b. Finally, we recap the main results and discuss future work in Section 7.

\section{Observations and Data Reduction} \label{sec:obs}
We observed five primary transits of WASP-43b with Gemini-S on February 29, 2023 (night 1), April 08, 2023 (night 2), April 26, 2023 (night 3), April 30, 2023 (night 4), and January 29, 2024 (night 5) as part of Program GS-2023A-Q-107 (PI Weiner Mansfield). The exposure time for all five nights was 85 seconds. Table \ref{table:obsparams} lists the total number of A-B pairs and orders removed for each observation.
\begin{table*}[ht]
\centering
\begin{tabular}{|c|c|c|c|c|} 
 \hline
 Observation & A-B pairs & $\phi$ & Average SNR & Orders removed \\ [1ex] 
 \hline
 Night 1 & 16 & -0.032 $<$ $\phi$ $<$ 0.033 & 125.9826 & 0-1, 20-26, 49-52 \\ 
 Night 2 & 15 & -0.035 $<$ $\phi$ $<$ 0.025 & 119.9661 & 0-1, 21-27, 51-53 \\
 Night 3 & 15 & -0.035 $<$ $\phi$ $<$ 0.022 & 127.4855 & 0-1, 21-26, 49-52 \\
 Night 4 & 14 & -0.027 $<$ $\phi$ $<$ 0.036 & 93.2027 & 0-2, 18, 20-27, 48-53 \\
 Night 5 & 16 & -0.027 $<$ $\phi$ $<$ 0.027 & 133.1136 & 0-1, 20-26, 49-52 \\ [1ex] 
 \hline
\end{tabular}
\caption{Observational parameters. Column 1 lists the observation, Column 2 lists the number of A-B pairs in transit, Column 3 lists the orbital phases spanned by the observation, Column 4 represents the average SNR prior to removing low-SNR orders, and Column 5 represents the orders discarded due to low transmittance and high telluric contamination.}
\label{table:obsparams}
\end{table*}

\begin{figure}[H]
  \centering
    \includegraphics[width=0.46\textwidth]{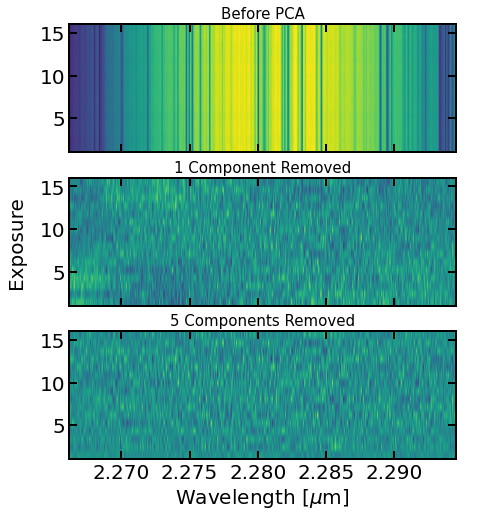}
    \caption{Progressive subtraction of SVs from the spectra obtained on night 5. Five components were removed by PCA, the optimal value for night 5. In the first panel, tellurics, stellar features, and the shape of the instrument throughput dominate the data. As components are progressively subtracted, the non-planetary signals are removed, and the planet's signal is hidden in the remaining noise. }
    \label{fig:pcanight5}
\end{figure}

We used the IGRINS Pipeline Package \cite[PLP,][]{Sim_2014, Lee_2016} to reduce and optimally extract the spectra and perform an initial wavelength calibration. We follow the methods of \cite{Line_2021} to separate the planet's signature from its host star and telluric contamination. First, we removed signal-to-noise orders with an SNR of less than 100; the orders removed for each night can be found in Table \ref{table:obsparams}. The average SNRs for each night before removing low SNR orders are included in Table \ref{table:obsparams}. Night 5 has the lowest airmass, resulting in the highest SNR. Humidity likely affects the average SNRs, although less so, as shown in Figure \ref{fig:humidity} in Section \ref{sec:crosscorr}. We then perform a secondary wavelength calibration to account for sub-pixel variations in wavelength. This calibration uses a stretch and shift to match the spectrum that was observed closest in time to the sky calibration observation. We assume that the spectrum measured directly before/after the sky calibration is exactly corrected to the “true" wavelengths. Finally, we use a singular value decomposition (SVD) to separate the planetary signal from non-planetary signals, specifically those from the host star or tellurics \citep{WM2024, wardenier2024phaseresolvingabsorptionsignatureswater}. We varied the number of singular vectors (SVs) removed from the data for each night to determine the optimal number needed to maximize the detection signal-to-noise ratio (SNR). Using Python's \texttt{numpy.linalg.svd} and varying the number of SVs removed from 1-8, we determined that removing 5 singular SVs from nights 1, 3, 4, and 5, and 6 SVs from night 2 best maximized the planetary signal. In testing beyond the number of SVs where the signal started to decrease, the decrease was consistent past the optimal number of SVs for the analysis. We tested whether optimizing the signal alone versus optimizing an injected signal \citep{2023Cheverall} affected the optimal number of SVs to remove. We found that the same number of SVs was preferred using both methods. The removal of telluric and stellar signals from the spectra obtained from night 5 can be seen in Figure \ref{fig:pcanight5}.
\section{Cross-Correlation } \label{sec:crosscorr}
\begin{figure}[H]
  \centering
    \includegraphics[width=0.46\textwidth]{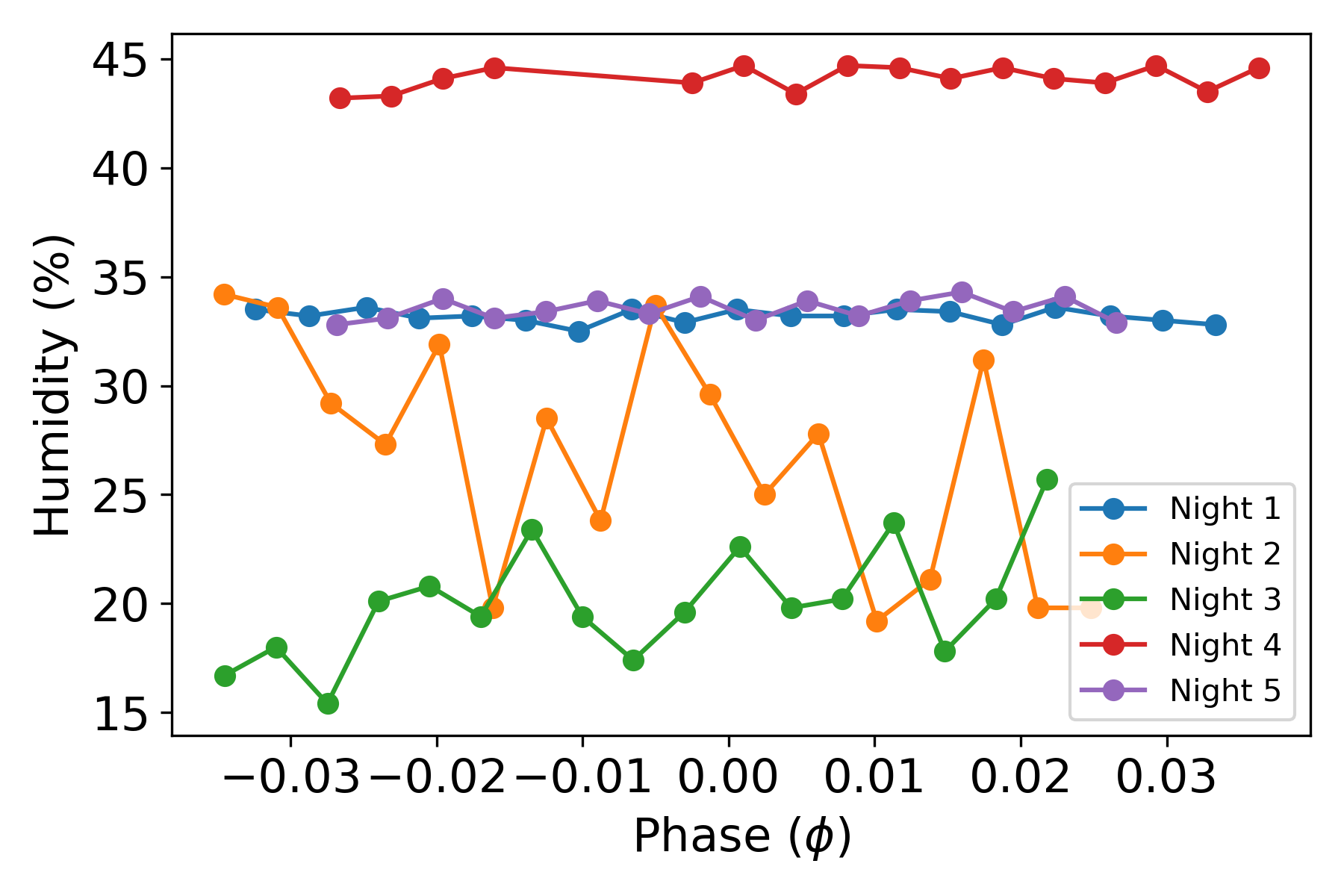}
    \caption{The humidities during each transit phase during night 1 (blue), night 2 (orange), night 3 (green), night 4 (red) and night 5 (purple). The higher humidity on night 4 resulted in a weaker detection of H$_2$O. Therefore, we excluded it from the analysis.}
    \label{fig:humidity}
\end{figure}

\begin{figure*}
  \centering
    \includegraphics[width=0.75\textwidth]{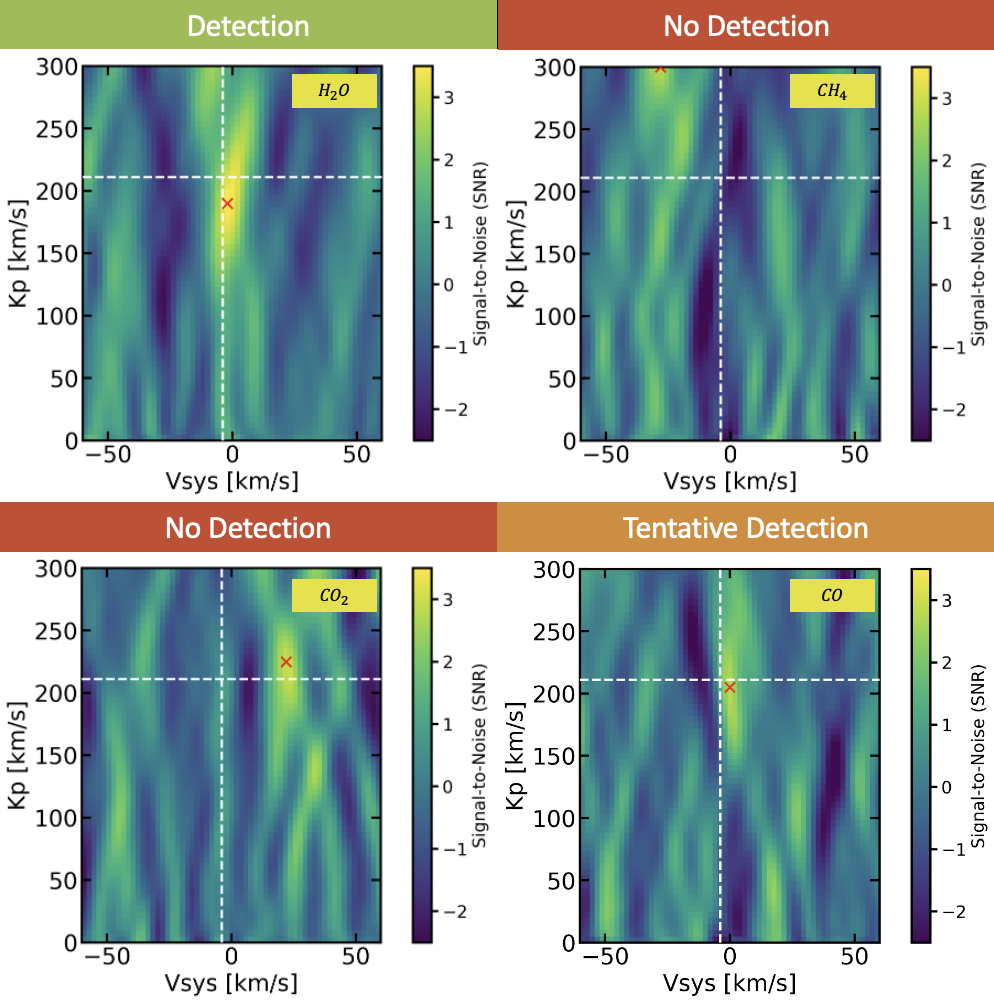}
    \caption{Cross-correlation SNRs as a function of systemic velocity (V$_{sys}$) and Keplerian velocity (K$_p$) for  H$_2$O, CH$_4$, CO$_2$, and CO  from the four nights combined. We detect H$_2$O at an SNR of 3.51 but do not detect any carbon-bearing species, although CO may be a tenative detection. The white dashed lines represent the V$_{sys}$ and K$_p$ values of WASP-43b from the literature \citep{Gaia2018, Bonomo2017}. The red x's indicate the maximum signal from the cross-correlation.}
    \label{fig:ccfs}
\end{figure*}

\begin{figure*}
  \centering
    \includegraphics[width=0.75\textwidth]{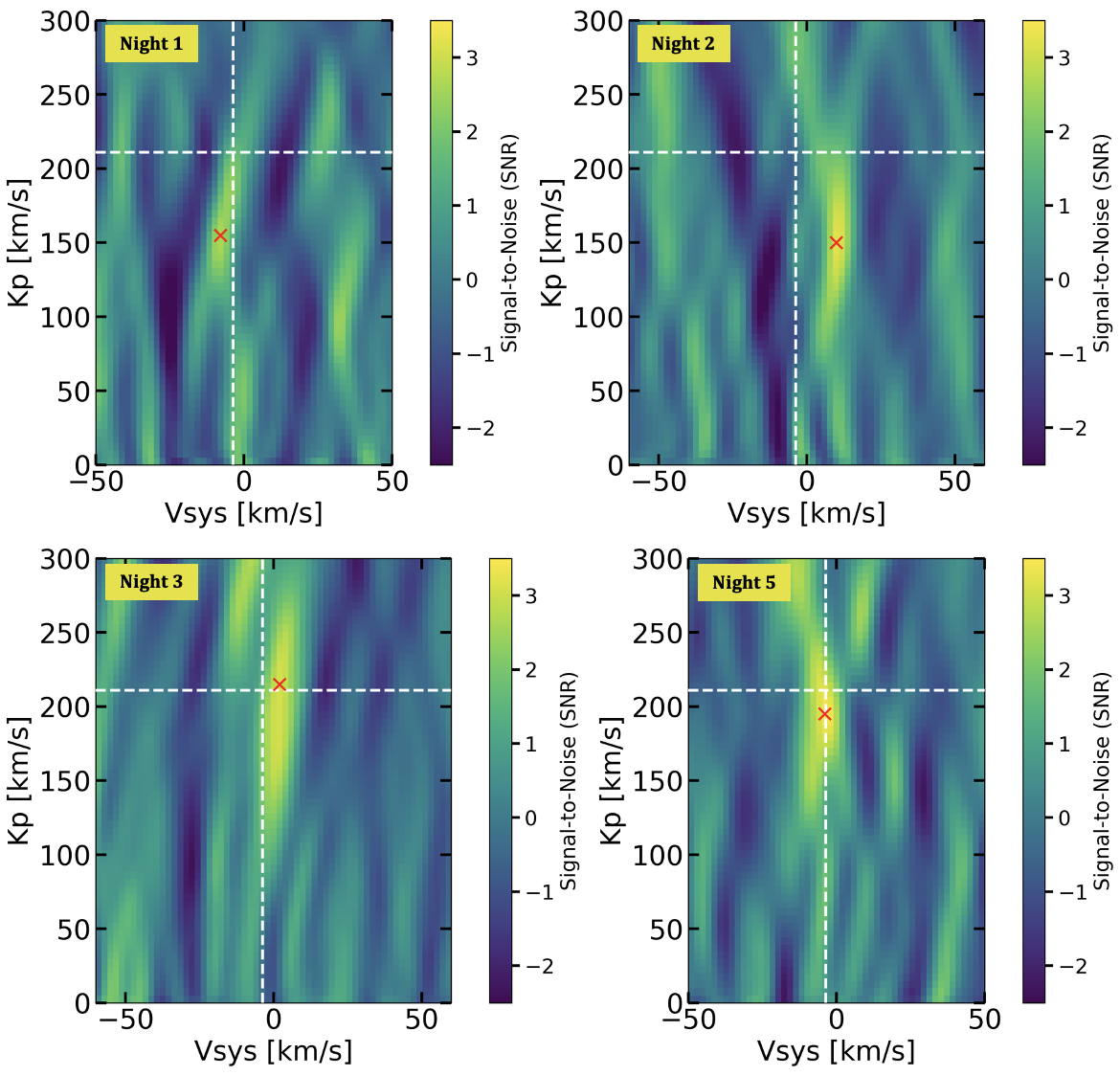}
    \caption{Cross-correlation SNRs as a function of systemic velocity (V$_{sys}$) and Keplerian velocity (K$_p$) for water, for Night 1, Night 2, Night 3, and Night 5. The white dashed lines represent the V$_{sys}$ and K$_p$ values of WASP-43b from the literature \citep{Gaia2018, Bonomo2017}. The red x's indicate the maximum signal from the cross-correlation.}
    \label{fig:waterccfs}
\end{figure*}

To determine which gases were detectable in our data, we cross-correlated our observations with a range of models. Following \citet{WM2024}, we used the ScCHIMERA framework to create a solar composition, radiative-convective-thermochemical equilibrium model for WASP-43b. We individually searched for single gases by modifying the model to only include H$_{2}$-He collisional-induced absorption and the single gas being searched for at a volume mixing ratio of $10^{-3}$. This volume mixing ratio was used because it allows for the species to be easily detected without being so abundant as to influence the bulk atmospheric composition. We used line lists for H$_{2}$O from POKAZATEL \citep{2018Polyansky, 2021ApJS..254...34G}, for CO and CH$_{4}$ from HITEMP \citep{2010HITEMP, methanehitemp, co_hitemp}, and for CO$_{2}$ from ExoMol \citep{Yurchenko2020}.\\
In running the CCF for the individual nights, we found that the results for night 4 showed no clear detection of water, and adding it to the cross-correlation did not improve the SNR of the detection. We determined that there was an increased amount of humidity on night 4 compared to the other nights. As a result, we excluded this observation from our combined CCF. The variation in humidity across the transit phases for each night can be seen in Figure \ref{fig:humidity}, highlighting the significant difference in humidity between night 4 and the other nights of observation. 
We combined four nights of data by performing the cross-correlation for each night individually and summing the cross-correlation strengths. We adopt the cross-correlation function (CCF) likelihood map from \cite{Brogi2019}. The cross-correlation strengths were converted to detection SNR by dividing the difference between the cross-correlation strengths and the 3$\sigma$ clipped median by the 3$\sigma$ clipped standard deviation. The median and standard deviation were calculated using \texttt{astropy.stats.sigma\_clipped\_stats}.  \\
Figure \ref{fig:ccfs} shows our results from the cross-correlation. To qualify as a detection, we required the maximum signal to have an SNR of $\geq$ 3.0 and appear near the literature values of  V$_{sys}$ and K$_p$. We detected H$_2$O with the combined data set using four nights, showing a detection SNR of 3.51. H$_2$O was also detected in the individual CCFs for nights 1, 2, 3, and 5, as seen in Figure~\ref{fig:waterccfs}. The detection SNR for H$_2$O for the individual nights are as follows: 2.66 (night 1), 3.22 (night 2), 3.09 (night 3), and 3.42 (night 5). We searched for CH$_4$, CO, and CO$_2$ but did not detect these molecules in the individual or combined CCFs.


For CO, we chose to run the cross-correlation for orders that contained the strongest absorption lines. In the 1.45 $-$ 2.45 $\mu$m wavelength range covered by IGRINS, the CO bands cover wavelengths between 1.56-1.64 $\mu$m and 2.29-2.45 $\mu$m. For nights 1, 2, and 5, we included only orders 0-1, which span wavelengths 2.29-2.42 $\mu$m and 20-26, which span wavelengths 1.55-1.60 $\mu$m. For night 3, we included only orders 0-1 and 21-27, which span wavelengths 1.53-1.59 $\mu$m. This resulted in a maximum cross-correlation signal with an SNR of 2.74 that appears close to the V$_{sys}$ and K$_p$ values of WASP-43b from the literature. Although this signal is weaker than the amplitude of the noise features, it is possible that this is a plausible signature of CO, which could be verified with additional data.   

 In Figure \ref{fig:ccfs}, the detection peak of H$_2$O appears offset relative to the expected values of $K_{p}$ and $V_{sys}$. The detection peak has a $K_{p}$ offset of $\Delta K_{p} = -21$ km/s and a $V_{sys}$ offset of $\Delta V_{sys} = +1.696$ km/s. In 3D hot Jupiter models developed by \cite{wardenier2023modelling}, they find that the offsets of the SNR peaks for $V_{sys}$ are a few km/s, whereas the peak offsets for $K_{p}$ can be more significant. Specifically, they find that H$_2$O can have velocity offsets of 5-20 km/s. The negative $K_{p}$ offset for H$_2$O is consistent with the findings of \cite{wardenier2023modelling}. However, our positive $V_{sys}$ offset is not consistent with the models of \cite{wardenier2023modelling}; their models show a negative $V_{sys}$ offset. A positive $V_{sys}$ offset was also observed for WASP-76b by \citet{WM2024}, who hypothesized that this offset could be due to magnetic effects. However, WASP-43 b is colder than WASP-76 b, so it may be less likely to have strong magnetic effects since it is not as strongly ionized. Another possible explanation for the $V_{sys}$ offset is clouds, which could impact the derived velocities, particularly if the limbs show different amounts of cloud coverage. However, as shown in Figure \ref{fig:retrieval}, the retrieval obtains error bars on $V_{sys}$ that are consistent with zero within 1$\sigma$. Future observations with a high enough SNR to study the phase-resolved $V_{sys}$ offset throughout the transit will be necessary to better understand the redshifted $V_{sys}$ for H$_2$O and to confirm the possible detection of a positive $V_{sys}$ offset.
 
\section{Retrievals} \label{sec:retrievals}
We used the retrieval framework of \citet{Line_2021} to estimate the abundance of water in the atmosphere of WASP-43b. We parameterized the atmosphere with an isothermal temperature ($T_{0}$, in K), a cloud-top pressure ($P_{c}$, in bars), and constant-with-altitude abundances for H$_{2}$O, CO, CO$_{2}$, and CH$_{4}$. We used an isothermal temperature because the relatively weak signal detected in these data precluded using a more complex T-P profile. Although the cross-correlation showed a strong detection of water only, we included the carbon-bearing species in our retrieval to allow us to place upper limits on their abundances and compare them to previous results. We used the line lists discussed in Section~\ref{sec:crosscorr} for H$_{2}$O, CO, CO$_{2}$, and CH$_{4}$, as well as line lists for H$_{2}$-H$_{2}$ and H$_{2}$-He collision-induced absorption \citep{Karman2019}. We also included as free parameters the stellar systemic velocity and the planet's orbital velocity ($V_{sys}$ and $K_{p}$) and a scale factor on the reference planet radius ($\times R_{p}$).

\begin{figure*}
    \centering
    \includegraphics[width=\linewidth]{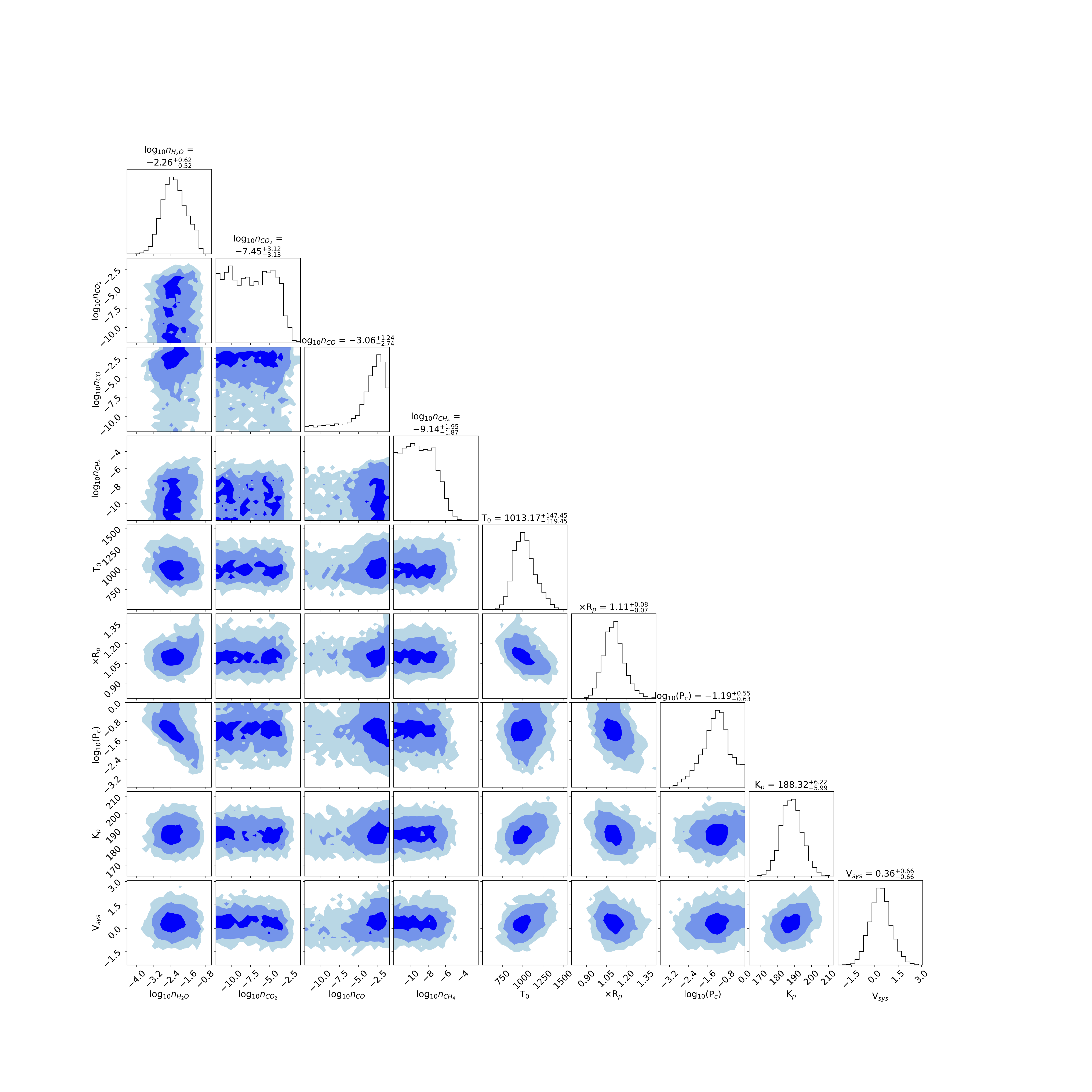}
    \caption{Posterior distributions from the retrieval on the atmospheric composition of WASP-43b. Off-diagonal plots show 2D posterior probabilities for pairs of parameters, with 1, 2, and $3\sigma$ intervals shaded in dark, medium, and light blue, respectively. On-diagonal plots show marginalized posterior probability distributions for each parameter. We find a constrained water abundance of $\log_{10}(\text{H}_2\text{O}) = -2.24^{+0.57}_{-0.48}$, an unconstrained CO abundance, and place $2\sigma$ upper limits on the abundances of CO$_{2}$ and CH$_{4}$ of $\log_{10}(\text{CO}_{2}) < -2.78$ and $\log_{10}(\text{CH}_4) < -5.89$.}
    \label{fig:retrieval}
\end{figure*}

We followed the method of \citet{WM2024} to convolve the model with kernels for instrumental and rotational broadening, shift it to appropriately account for the barycentric, stellar, and planetary velocities, and perform Bayesian inference using the log-likelihood framework of \citet{Brogi2019}. The parameter estimation was performed using Nested Sampling and the \texttt{pymultinest} package \citep{Buchnerpymulti}. Figure~\ref{fig:retrieval} shows the posterior distributions derived from the retrieval. The retrieved water abundance was $\log_{10}(\text{H}_2\text{O}) = -2.24^{+0.57}_{-0.48}$, while the CO abundance was essentially unconstrained. We retrieved upper limits on the CO$_{2}$ and CH$_{4}$ abundances of $\log_{10}(\text{CO}_{2}) < -2.78$ and $\log_{10}(\text{CH}_4) < -5.89$. From the retrieved water abundance and upper limits on the carbon-bearing species, we calculate a 2$\sigma$ upper limit of C/O $<$ 0.95.

\section{Discussion} \label{sec:discussion}
Several other studies have measured the water abundance in the atmosphere of WASP-43b. Table \ref{table:mixingratios} shows retrieved water volume mixing ratios obtained from other analyses, both low- and high-resolution, of the atmosphere of WASP-43b and how they compare to this work. \\
\begin{table*}[ht]
\centering
\begin{tabular}{|c|c|c|c|} 
 \hline
 $\log_{10}(\text{H}_2\text{O})$ Volume Mixing Ratio & Observatory/Instrument & Reference & Spectra Type \\ [1ex] 
 \hline
 $-2.24^{+0.57}_{-0.48}$ & Gemini-S/IGRINS & This work & Transmission\\ 
 $-3.6^{+0.8}_{-0.9}$ & \textit{HST}/WFC3 & \cite{Kreidberg_2014} & Transmission\\
 $-2.9^{+0.6}_{-0.6}$ & \textit{HST}/WFC3 & \cite{Kreidberg_2014} & Emission\\
 $-4.36^{+2.10}_{-2.10}$ & \textit{HST}/WFC3 & \cite{Tsiaras2018} & Transmission \\
 $-3.68^{+0.92}_{-0.88}$ & \textit{HST}/\textit{Spitzer} & \cite{2019Welbanks} & Transmission\\
 $-2.78^{+1.38}_{-1.47}$ & Magellan/IMACS, \textit{HST}/WFC3 & \cite{2020Weaver} & Emission\\
 $-4.4^{+0.3}_{-0.3}$ & VLT/CRIRES$^+$ & \cite{2023Lesjak} & Emission \\
 $-3.86^{+0.69}_{-0.36}$ & \textit{JWST}/MIRI & \cite{Yang2024} & Emission\\[1ex] 
 \hline
\end{tabular}
\caption{Water volume mixing ratios for WASP-43b, measured through direct observations. \cite{2023Lesjak} and \cite{Yang2024} obtained their results from the analysis of emission spectra of WASP-43b. This work, \cite{Tsiaras2018}, \cite{2019Welbanks}, and \cite{2020Weaver} use transmission spectra to obtain the results. \cite{Kreidberg_2014} uses both emission and transmission spectra to analyze the atmosphere of WASP-43b.}
\label{table:mixingratios}
\end{table*}
\begin{figure*}
  \centering
    \includegraphics[width=\linewidth]{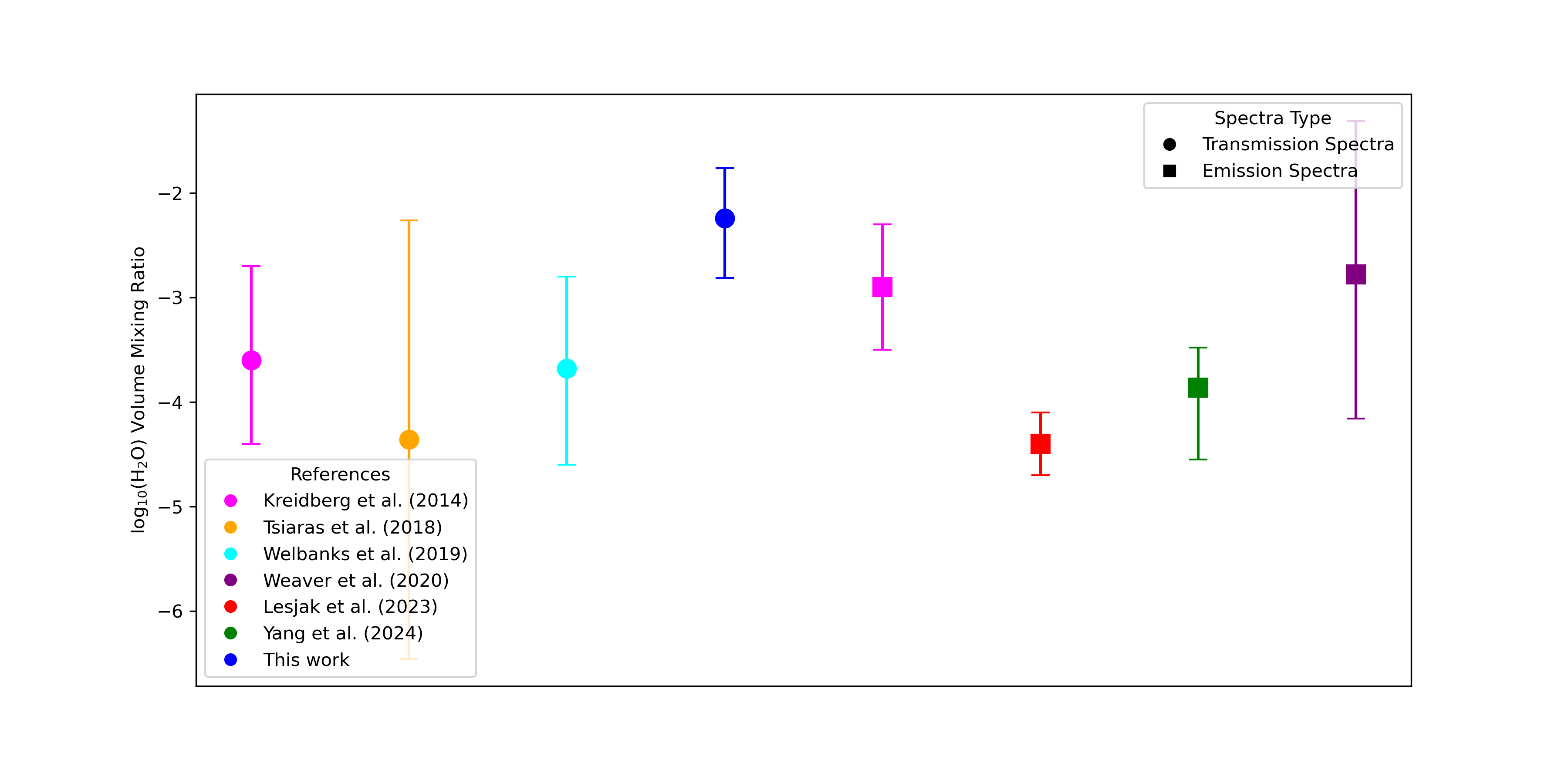}
    \caption{Water volume mixing ratios for WASP-43b, measured through direct observations. The circular points indicate that the measurement was from a transmission spectrum, whereas the square points indicate that the measurement was from emission spectra. References are listed in the legend in the bottom left corner and in Table \ref{table:mixingratios}.}
    \label{fig:tab2}
\end{figure*}
Our findings align with those of \cite{Kreidberg_2014}. Using the WFC3 instrument on \textit{HST}, they detected water in the atmosphere of WASP-43b from the emission spectrum at 11.7$\sigma$ confidence and 5$\sigma$ confidence from the transmission spectrum. They combine these observations to constrain the water volume mixing ratio to 2.4 $\times$ 10$^{-4}$ - 2.1 $\times$ 10$^{-3}$ at 1$\sigma$ confidence. Our constraint on the water volume mixing ratio of $\log_{10}(\text{H}_2\text{O}) = -2.24^{+0.57}_{-0.48}$ agrees within 1$\sigma$. \cite{Tsiaras2018} performed an analysis of the \textit{HST}/WFC3 data and obtained an abundance of $\log_{10}(\text{H}_2\text{O}) = -4.36^{+2.10}_{-2.10}$. Our value for the water volume mixing ratio is in better agreement with that of \citep{Kreidberg_2014}; differences in the modeling/retrievals done in \cite{Kreidberg_2014} and \cite{Tsiaras2018} may explain the disagreement between their abundances.\\
\cite{2023Lesjak} obtained the first high-resolution emission spectra of WASP-43b with the CRIRES$^+$ spectrograph. They detected water and retrieved an abundance of $\log_{10}(\text{H}_2\text{O}) = -4.4\pm0.03$, which is lower than our water abundance. Differences between the water abundances are likely a result of observing differences. For instance, the CRIRES$^+$ results come from dayside observations of WASP-43b, which view different parts of the planet compared to transit observations. Emission spectroscopy probes the dayside of the planet, whereas transmission spectra observe the planet's limbs. Transmission spectra are sensitive to the effects of clouds at the planet's limb, while there are likely no clouds on the planet's dayside. 
The phase curve of WASP-43b from \textit{JWST} suggests a clear dayside and a cloudy nightside \citep{Bell2024}. The corner plot presents a mild degeneracy between the water abundance and cloud top pressure, which shows how clouds can cause differences between emission and transmission spectra of the same planet. \cite{2023Lesjak} also detect CO in the atmosphere at an abundance of $\log_{10}(\text{CO}) = -3.0^{+0.8}_{-0.6}$. 
The abundance of CO retrieved from CRIRES$^+$ is consistent with the peak of our retrieved CO posterior, so if our tentative detection is confirmed, our findings would agree.
The non-detections of CH$_4$ or CO$_2$ from the CRIRES$^+$ results also agree with our findings. Additionally, our upper limit on the C/O ratio of C/O$<0.95$ agrees with the C/O ratio of $0.78\pm0.09$ measured by \citet{2023Lesjak}.\\
The detection of water is also consistent with recent findings from \textit{JWST} \citep{Bell2024, Yang2024}. A full orbit of WASP-43b was observed in the 5–12 $\mu$m range with \textit{JWST} MIRI in Low Resolution Spectroscopy slitless mode. From the emission spectra, \cite{Bell2024} detected water vapor at a confidence of $\sim$ 3-4$\sigma$. \cite{Yang2024} use a parametric 2D model to measure the global water abundance and retrieve an abundance of $\log_{10}(\text{H}_2\text{O}) = -3.86^{+0.69}_{-0.36}$. This abundance agrees with the CRIRES$^+$ results and is consistent within 2$\sigma$ of our result. Differences between our results and the \textit{JWST}/MIRI observations could be because our observations utilize the transmission spectrum, which is localized to the high-altitude limb, whereas the 2D model used by \citet{Yang2024} retrieves a global abundance. \cite{Yang2024} also detect CO in the atmosphere of WASP-43b at an abundance of $\log_{10}(\text{CO}) = -3.25^{+0.58}_{-0.46}$, as well as weak evidence of C$\text{O}_2$ with an abundance of $\log_{10}(\text{CO}_{2}) = -4.60^{+0.55}_{-0.50}$. The abundance of C$\text{O}_2$ from \cite{Yang2024} agrees with the upper limit of $\log_{10}(\text{CO}_{2}) < -2.78$ that we retrieved. Furthermore, their abundance of CO also agrees with the peak of our retrieved CO posterior. 
If our tentative CO detection is revealed to be a true detection, then there would be a stronger agreement between our results and those of \textit{JWST}. \cite{Yang2024} report a non-detection of CH$_4$, agreeing with our findings.
\section{Conclusions} \label{sec:style}
We present observations of the hot Jupiter WASP-43b observed by IGRINS in the 1.45 $-$ 2.45 $\mu$m wavelength range. 
We detect the presence of $\text{H}_2$O at a detection SNR of 3.51 and retrieved a water volume mixing ratio of $\log_{10}(\text{H}_2\text{O}) = -2.24^{+0.57}_{-0.48}$. Our detection of H$_2$O and retrieved water volume mixing ratio are consistent with previous analyses of transmission spectra from \textit{HST} and \textit{JWST} \citep{Kreidberg_2014, Bell2024,Yang2024}. We do not detect the presence of C$\text{O}_2$ and C$\text{H}_4$ and get a weak suggestion of CO. The retrieved median abundance of CO agrees with the CO abundances of \cite{2023Lesjak} and \cite{Yang2024}. We obtain a 2$\sigma$ upper limit on the C/O ratio of C/O $<$ 0.95. Future high-resolution observations of WASP-43b, especially by the upcoming Extremely Large Telescopes \citep{2022SPIE12182E..1CF, 2023ConPh..64...47P}, will be useful to better constrain the abundances of carbon- and oxygen-bearing species to get a more accurate C/O ratio.

\begin{acknowledgements}

We thank the anonymous referee for their helpful feedback. Based on observations obtained as part of Program IDs GS-2023A-Q-107 at the international Gemini Observatory, a program of NSF's NOIRLab, which is managed by the Association of Universities for Research in Astronomy (AURA) under a cooperative agreement with the National Science Foundation on behalf of the Gemini Observatory partnership: the National Science Foundation (United States), National Research Council (Canada), Agencia Nacional de Investigaci\'{o}n y Desarrollo (Chile), Ministerio de Ciencia, Tecnolog\'{i}a e Innovaci\'{o}n (Argentina), Minist\'{e}rio da Ci\^{e}ncia, Tecnologia, Inova\c{c}\~{o}es e Comunica\c{c}\~{o}es (Brazil), and Korea Astronomy and Space Science Institute (Republic of Korea). This work used the Immersion Grating Infrared Spectrometer (IGRINS) that was developed under a collaboration between the University of Texas at Austin and the Korea Astronomy and Space Science Institute (KASI) with the financial support of the Mt. Cuba Astronomical Foundation, of the US National Science Foundation under grants AST-1229522 and AST-1702267, of the McDonald Observatory of the University of Texas at Austin, of the Korean GMT Project of KASI, and Gemini Observatory. M.W.M. acknowledges support from the NASA Hubble Fellowship grant HST-HF2-51485.001-A awarded by the Space Telescope Science Institute, which is operated by AURA, Inc., for NASA, under contract NAS5-26555. The authors acknowledge Research Computing at Arizona State University for providing high-performance computing resources that have contributed to the research results reported within this paper.

\end{acknowledgements}

\clearpage
\bibliography{main}

\begin{thebibliography}{}
\expandafter\ifx\csname natexlab\endcsname\relax\def\natexlab#1{#1}\fi
\providecommand{\url}[1]{\href{#1}{#1}}
\providecommand{\dodoi}[1]{doi:~\href{http://doi.org/#1}{\nolinkurl{#1}}}
\providecommand{\doeprint}[1]{\href{http://ascl.net/#1}{\nolinkurl{http://ascl.net/#1}}}
\providecommand{\doarXiv}[1]{\href{https://arxiv.org/abs/#1}{\nolinkurl{https://arxiv.org/abs/#1}}}

\bibitem[{{Bell} {et~al.}(2024){Bell}, {Crouzet}, {Cubillos}, {Kreidberg}, {Piette}, {Roman}, {Barstow}, {Blecic}, {Carone}, {Coulombe}, {Ducrot}, {Hammond}, {Mendon{\c{c}}a}, {Moses}, {Parmentier}, {Stevenson}, {Teinturier}, {Zhang}, {Batalha}, {Bean}, {Benneke}, {Charnay}, {Chubb}, {Demory}, {Gao}, {Lee}, {L{\'o}pez-Morales}, {Morello}, {Rauscher}, {Sing}, {Tan}, {Venot}, {Wakeford}, {Aggarwal}, {Ahrer}, {Alam}, {Baeyens}, {Barrado}, {Caceres}, {Carter}, {Casewell}, {Challener}, {Crossfield}, {Decin}, {D{\'e}sert}, {Dobbs-Dixon}, {Dyrek}, {Espinoza}, {Feinstein}, {Gibson}, {Harrington}, {Helling}, {Hu}, {Iro}, {Kempton}, {Kendrew}, {Komacek}, {Krick}, {Lagage}, {Leconte}, {Lendl}, {Lewis}, {Lothringer}, {Malsky}, {Mancini}, {Mansfield}, {Mayne}, {Evans-Soma}, {Molaverdikhani}, {Nikolov}, {Nixon}, {Palle}, {Petit dit de la Roche}, {Piaulet}, {Powell}, {Rackham}, {Schneider}, {Steinrueck}, {Taylor}, {Welbanks}, {Yurchenko}, {Zhang}, \& {Zieba}}]{Bell2024}
{Bell}, T.~J., {Crouzet}, N., {Cubillos}, P.~E., {et~al.} 2024, Nature Astronomy, \dodoi{10.1038/s41550-024-02230-x}

\bibitem[{Birkby(2018)}]{birkby2018exoplanetatmosphereshighspectral}
Birkby, J.~L. 2018, Exoplanet Atmospheres at High Spectral Resolution.
\newblock \doarXiv{1806.04617}

\bibitem[{{Blecic} {et~al.}(2014){Blecic}, {Harrington}, {Madhusudhan}, {Stevenson}, {Hardy}, {Cubillos}, {Hardin}, {Bowman}, {Nymeyer}, {Anderson}, {Hellier}, {Smith}, \& {Collier Cameron}}]{2014Blecic}
{Blecic}, J., {Harrington}, J., {Madhusudhan}, N., {et~al.} 2014, \apj, 781, 116, \dodoi{10.1088/0004-637X/781/2/116}

\bibitem[{{Bonomo} {et~al.}(2017){Bonomo}, {Desidera}, {Benatti}, {Borsa}, {Crespi}, {Damasso}, {Lanza}, {Sozzetti}, {Lodato}, {Marzari}, {Boccato}, {Claudi}, {Cosentino}, {Covino}, {Gratton}, {Maggio}, {Micela}, {Molinari}, {Pagano}, {Piotto}, {Poretti}, {Smareglia}, {Affer}, {Biazzo}, {Bignamini}, {Esposito}, {Giacobbe}, {H{\'e}brard}, {Malavolta}, {Maldonado}, {Mancini}, {Martinez Fiorenzano}, {Masiero}, {Nascimbeni}, {Pedani}, {Rainer}, \& {Scandariato}}]{Bonomo2017}
{Bonomo}, A.~S., {Desidera}, S., {Benatti}, S., {et~al.} 2017, \aap, 602, A107, \dodoi{10.1051/0004-6361/201629882}

\bibitem[{Brogi \& Line(2019)}]{Brogi2019}
Brogi, M., \& Line, M.~R. 2019, The Astronomical Journal, 157, 114, \dodoi{10.3847/1538-3881/aaffd3}

\bibitem[{{Brogi} {et~al.}(2023){Brogi}, {Emeka-Okafor}, {Line}, {Gandhi}, {Pino}, {Kempton}, {Rauscher}, {Parmentier}, {Bean}, {Mace}, {Cowan}, {Shkolnik}, {Wardenier}, {Mansfield}, {Welbanks}, {Smith}, {Fortney}, {Birkby}, {Zalesky}, {Dang}, {Patience}, \& {D{\'e}sert}}]{Brogi2023}
{Brogi}, M., {Emeka-Okafor}, V., {Line}, M.~R., {et~al.} 2023, \aj, 165, 91, \dodoi{10.3847/1538-3881/acaf5c}

\bibitem[{{Buchner} {et~al.}(2014){Buchner}, {Georgakakis}, {Nandra}, {Hsu}, {Rangel}, {Brightman}, {Merloni}, {Salvato}, {Donley}, \& {Kocevski}}]{Buchnerpymulti}
{Buchner}, J., {Georgakakis}, A., {Nandra}, K., {et~al.} 2014, \aap, 564, A125, \dodoi{10.1051/0004-6361/201322971}

\bibitem[{Chachan {et~al.}(2023)Chachan, Knutson, Lothringer, \& Blake}]{Chachan_2023}
Chachan, Y., Knutson, H.~A., Lothringer, J., \& Blake, G.~A. 2023, The Astrophysical Journal, 943, 112, \dodoi{10.3847/1538-4357/aca614}

\bibitem[{{Cheverall} {et~al.}(2023){Cheverall}, {Madhusudhan}, \& {Holmberg}}]{2023Cheverall}
{Cheverall}, C.~J., {Madhusudhan}, N., \& {Holmberg}, M. 2023, \mnras, 522, 661, \dodoi{10.1093/mnras/stad648}

\bibitem[{Deibert {et~al.}(2023)Deibert, de~Mooij, Jayawardhana, Turner, Ridden-Harper, Hood, Fortney, Flagg, Fossati, Allart, Brogi, \& MacDonald}]{deibert2023exogems}
Deibert, E.~K., de~Mooij, E. J.~W., Jayawardhana, R., {et~al.} 2023, ExoGemS High-Resolution Transmission Spectroscopy of WASP-76b with GRACES.
\newblock \doarXiv{2307.16738}

\bibitem[{{Fanson} {et~al.}(2022){Fanson}, {Bernstein}, {Ashby}, {Bigelow}, {Brossus}, {Burgett}, {Demers}, {Fischer}, {Figueroa}, {Groark}, {Laskin}, {Millan-Gabet}, {Park}, {Pi}, {Turner}, \& {Walls}}]{2022SPIE12182E..1CF}
{Fanson}, J., {Bernstein}, R., {Ashby}, D., {et~al.} 2022, in Society of Photo-Optical Instrumentation Engineers (SPIE) Conference Series, Vol. 12182, Ground-based and Airborne Telescopes IX, ed. H.~K. {Marshall}, J.~{Spyromilio}, \& T.~{Usuda}, 121821C, \dodoi{10.1117/12.2631694}

\bibitem[{{Gaia Collaboration} {et~al.}(2018){Gaia Collaboration}, {Brown}, {Vallenari}, {Prusti}, {de Bruijne}, {Babusiaux}, {Bailer-Jones}, {Biermann}, {Evans}, {Eyer}, {Jansen}, {Jordi}, {Klioner}, {Lammers}, {Lindegren}, {Luri}, {Mignard}, {Panem}, {Pourbaix}, {Randich}, {Sartoretti}, {Siddiqui}, {Soubiran}, {van Leeuwen}, {Walton}, {Arenou}, {Bastian}, {Cropper}, {Drimmel}, {Katz}, {Lattanzi}, {Bakker}, {Cacciari}, {Casta{\~n}eda}, {Chaoul}, {Cheek}, {De Angeli}, {Fabricius}, {Guerra}, {Holl}, {Masana}, {Messineo}, {Mowlavi}, {Nienartowicz}, {Panuzzo}, {Portell}, {Riello}, {Seabroke}, {Tanga}, {Th{\'e}venin}, {Gracia-Abril}, {Comoretto}, {Garcia-Reinaldos}, {Teyssier}, {Altmann}, {Andrae}, {Audard}, {Bellas-Velidis}, {Benson}, {Berthier}, {Blomme}, {Burgess}, {Busso}, {Carry}, {Cellino}, {Clementini}, {Clotet}, {Creevey}, {Davidson}, {De Ridder}, {Delchambre}, {Dell'Oro}, {Ducourant}, {Fern{\'a}ndez-Hern{\'a}ndez}, {Fouesneau}, {Fr{\'e}mat}, {Galluccio}, {Garc{\'\i}a-Torres},
  {Gonz{\'a}lez-N{\'u}{\~n}ez}, {Gonz{\'a}lez-Vidal}, {Gosset}, {Guy}, {Halbwachs}, {Hambly}, {Harrison}, {Hern{\'a}ndez}, {Hestroffer}, {Hodgkin}, {Hutton}, {Jasniewicz}, {Jean-Antoine-Piccolo}, {Jordan}, {Korn}, {Krone-Martins}, {Lanzafame}, {Lebzelter}, {L{\"o}ffler}, {Manteiga}, {Marrese}, {Mart{\'\i}n-Fleitas}, {Moitinho}, {Mora}, {Muinonen}, {Osinde}, {Pancino}, {Pauwels}, {Petit}, {Recio-Blanco}, {Richards}, {Rimoldini}, {Robin}, {Sarro}, {Siopis}, {Smith}, {Sozzetti}, {S{\"u}veges}, {Torra}, {van Reeven}, {Abbas}, {Abreu Aramburu}, {Accart}, {Aerts}, {Altavilla}, {{\'A}lvarez}, {Alvarez}, {Alves}, {Anderson}, {Andrei}, {Anglada Varela}, {Antiche}, {Antoja}, {Arcay}, {Astraatmadja}, {Bach}, {Baker}, {Balaguer-N{\'u}{\~n}ez}, {Balm}, {Barache}, {Barata}, {Barbato}, {Barblan}, {Barklem}, {Barrado}, {Barros}, {Barstow}, {Bartholom{\'e} Mu{\~n}oz}, {Bassilana}, {Becciani}, {Bellazzini}, {Berihuete}, {Bertone}, {Bianchi}, {Bienaym{\'e}}, {Blanco-Cuaresma}, {Boch}, {Boeche}, {Bombrun}, {Borrachero},
  {Bossini}, {Bouquillon}, {Bourda}, {Bragaglia}, {Bramante}, {Breddels}, {Bressan}, {Brouillet}, {Br{\"u}semeister}, {Brugaletta}, {Bucciarelli}, {Burlacu}, {Busonero}, {Butkevich}, {Buzzi}, {Caffau}, {Cancelliere}, {Cannizzaro}, {Cantat-Gaudin}, {Carballo}, {Carlucci}, {Carrasco}, {Casamiquela}, {Castellani}, {Castro-Ginard}, {Charlot}, {Chemin}, {Chiavassa}, {Cocozza}, {Costigan}, {Cowell}, {Crifo}, {Crosta}, {Crowley}, {Cuypers}, {Dafonte}, {Damerdji}, {Dapergolas}, {David}, {David}, {de Laverny}, {De Luise}, {De March}, {de Martino}, {de Souza}, {de Torres}, {Debosscher}, {del Pozo}, {Delbo}, {Delgado}, {Delgado}, {Di Matteo}, {Diakite}, {Diener}, {Distefano}, {Dolding}, {Drazinos}, {Dur{\'a}n}, {Edvardsson}, {Enke}, {Eriksson}, {Esquej}, {Eynard Bontemps}, {Fabre}, {Fabrizio}, {Faigler}, {Falc{\~a}o}, {Farr{\`a}s Casas}, {Federici}, {Fedorets}, {Fernique}, {Figueras}, {Filippi}, {Findeisen}, {Fonti}, {Fraile}, {Fraser}, {Fr{\'e}zouls}, {Gai}, {Galleti}, {Garabato}, {Garc{\'\i}a-Sedano}, {Garofalo},
  {Garralda}, {Gavel}, {Gavras}, {Gerssen}, {Geyer}, {Giacobbe}, {Gilmore}, {Girona}, {Giuffrida}, {Glass}, {Gomes}, {Granvik}, {Gueguen}, {Guerrier}, {Guiraud}, {Guti{\'e}rrez-S{\'a}nchez}, {Haigron}, {Hatzidimitriou}, {Hauser}, {Haywood}, {Heiter}, {Helmi}, {Heu}, {Hilger}, {Hobbs}, {Hofmann}, {Holland}, {Huckle}, {Hypki}, {Icardi}, {Jan{\ss}en}, {Jevardat de Fombelle}, {Jonker}, {Juh{\'a}sz}, {Julbe}, {Karampelas}, {Kewley}, {Klar}, {Kochoska}, {Kohley}, {Kolenberg}, {Kontizas}, {Kontizas}, {Koposov}, {Kordopatis}, {Kostrzewa-Rutkowska}, {Koubsky}, {Lambert}, {Lanza}, {Lasne}, {Lavigne}, {Le Fustec}, {Le Poncin-Lafitte}, {Lebreton}, {Leccia}, {Leclerc}, {Lecoeur-Taibi}, {Lenhardt}, {Leroux}, {Liao}, {Licata}, {Lindstr{\o}m}, {Lister}, {Livanou}, {Lobel}, {L{\'o}pez}, {Managau}, {Mann}, {Mantelet}, {Marchal}, {Marchant}, {Marconi}, {Marinoni}, {Marschalk{\'o}}, {Marshall}, {Martino}, {Marton}, {Mary}, {Massari}, {Matijevi{\v{c}}}, {Mazeh}, {McMillan}, {Messina}, {Michalik}, {Millar}, {Molina}, {Molinaro},
  {Moln{\'a}r}, {Montegriffo}, {Mor}, {Morbidelli}, {Morel}, {Morris}, {Mulone}, {Muraveva}, {Musella}, {Nelemans}, {Nicastro}, {Noval}, {O'Mullane}, {Ord{\'e}novic}, {Ord{\'o}{\~n}ez-Blanco}, {Osborne}, {Pagani}, {Pagano}, {Pailler}, {Palacin}, {Palaversa}, {Panahi}, {Pawlak}, {Piersimoni}, {Pineau}, {Plachy}, {Plum}, {Poggio}, {Poujoulet}, {Pr{\v{s}}a}, {Pulone}, {Racero}, {Ragaini}, {Rambaux}, {Ramos-Lerate}, {Regibo}, {Reyl{\'e}}, {Riclet}, {Ripepi}, {Riva}, {Rivard}, {Rixon}, {Roegiers}, {Roelens}, {Romero-G{\'o}mez}, {Rowell}, {Royer}, {Ruiz-Dern}, {Sadowski}, {Sagrist{\`a} Sell{\'e}s}, {Sahlmann}, {Salgado}, {Salguero}, {Sanna}, {Santana-Ros}, {Sarasso}, {Savietto}, {Schultheis}, {Sciacca}, {Segol}, {Segovia}, {S{\'e}gransan}, {Shih}, {Siltala}, {Silva}, {Smart}, {Smith}, {Solano}, {Solitro}, {Sordo}, {Soria Nieto}, {Souchay}, {Spagna}, {Spoto}, {Stampa}, {Steele}, {Steidelm{\"u}ller}, {Stephenson}, {Stoev}, {Suess}, {Surdej}, {Szabados}, {Szegedi-Elek}, {Tapiador}, {Taris}, {Tauran}, {Taylor},
  {Teixeira}, {Terrett}, {Teyssandier}, {Thuillot}, {Titarenko}, {Torra Clotet}, {Turon}, {Ulla}, {Utrilla}, {Uzzi}, {Vaillant}, {Valentini}, {Valette}, {van Elteren}, {Van Hemelryck}, {van Leeuwen}, {Vaschetto}, {Vecchiato}, {Veljanoski}, {Viala}, {Vicente}, {Vogt}, {von Essen}, {Voss}, {Votruba}, {Voutsinas}, {Walmsley}, {Weiler}, {Wertz}, {Wevers}, {Wyrzykowski}, {Yoldas}, {{\v{Z}}erjal}, {Ziaeepour}, {Zorec}, {Zschocke}, {Zucker}, {Zurbach}, \& {Zwitter}}]{Gaia2018}
{Gaia Collaboration}, {Brown}, A.~G.~A., {Vallenari}, A., {et~al.} 2018, \aap, 616, A1, \dodoi{10.1051/0004-6361/201833051}

\bibitem[{{Gandhi} {et~al.}(2019){Gandhi}, {Madhusudhan}, {Hawker}, \& {Piette}}]{Gandhi2019}
{Gandhi}, S., {Madhusudhan}, N., {Hawker}, G., \& {Piette}, A. 2019, \aj, 158, 228, \dodoi{10.3847/1538-3881/ab4efc}

\bibitem[{{Gharib-Nezhad} {et~al.}(2021){Gharib-Nezhad}, {Iyer}, {Line}, {Freedman}, {Marley}, \& {Batalha}}]{2021ApJS..254...34G}
{Gharib-Nezhad}, E., {Iyer}, A.~R., {Line}, M.~R., {et~al.} 2021, \apjs, 254, 34, \dodoi{10.3847/1538-4365/abf504}

\bibitem[{{Hargreaves} {et~al.}(2020){Hargreaves}, {Gordon}, {Rey}, {Nikitin}, {Tyuterev}, {Kochanov}, \& {Rothman}}]{methanehitemp}
{Hargreaves}, R.~J., {Gordon}, I.~E., {Rey}, M., {et~al.} 2020, \apjs, 247, 55, \dodoi{10.3847/1538-4365/ab7a1a}

\bibitem[{{Hellier} {et~al.}(2011){Hellier}, {Anderson}, {Collier Cameron}, {Gillon}, {Jehin}, {Lendl}, {Maxted}, {Pepe}, {Pollacco}, {Queloz}, {S{\'e}gransan}, {Smalley}, {Smith}, {Southworth}, {Triaud}, {Udry}, \& {West}}]{2011Hellier}
{Hellier}, C., {Anderson}, D.~R., {Collier Cameron}, A., {et~al.} 2011, \aap, 535, L7, \dodoi{10.1051/0004-6361/201117081}

\bibitem[{{Irwin} {et~al.}(2020){Irwin}, {Parmentier}, {Taylor}, {Barstow}, {Aigrain}, {Lee}, \& {Garland}}]{2020MNRAS.493..106I}
{Irwin}, P. G.~J., {Parmentier}, V., {Taylor}, J., {et~al.} 2020, \mnras, 493, 106, \dodoi{10.1093/mnras/staa238}

\bibitem[{{Karman} {et~al.}(2019){Karman}, {Gordon}, {van der Avoird}, {Baranov}, {Boulet}, {Drouin}, {Groenenboom}, {Gustafsson}, {Hartmann}, {Kurucz}, {Rothman}, {Sun}, {Sung}, {Thalman}, {Tran}, {Wishnow}, {Wordsworth}, {Vigasin}, {Volkamer}, \& {van der Zande}}]{Karman2019}
{Karman}, T., {Gordon}, I.~E., {van der Avoird}, A., {et~al.} 2019, \icarus, 328, 160, \dodoi{10.1016/j.icarus.2019.02.034}

\bibitem[{Kreidberg {et~al.}(2014)Kreidberg, Bean, Désert, Line, Fortney, Madhusudhan, Stevenson, Showman, Charbonneau, McCullough, Seager, Burrows, Henry, Williamson, Kataria, \& Homeier}]{Kreidberg_2014}
Kreidberg, L., Bean, J.~L., Désert, J.-M., {et~al.} 2014, The Astrophysical Journal, 793, L27, \dodoi{10.1088/2041-8205/793/2/l27}

\bibitem[{Lee \& Gullikson(2016)}]{Lee_2016}
Lee, J.-J., \& Gullikson, K. 2016, plp: v2.1 alpha 3, v2.1-alpha.3,  Zenodo, \dodoi{10.5281/zenodo.56067}

\bibitem[{{Lesjak} {et~al.}(2023){Lesjak}, {Nortmann}, {Yan}, {Cont}, {Reiners}, {Piskunov}, {Hatzes}, {Boldt-Christmas}, {Czesla}, {Heiter}, {Kochukhov}, {Lavail}, {Nagel}, {Rains}, {Rengel}, {Rodler}, {Seemann}, \& {Shulyak}}]{2023Lesjak}
{Lesjak}, F., {Nortmann}, L., {Yan}, F., {et~al.} 2023, \aap, 678, A23, \dodoi{10.1051/0004-6361/202347151}

\bibitem[{{Li} {et~al.}(2015){Li}, {Gordon}, {Rothman}, {Tan}, {Hu}, {Kassi}, {Campargue}, \& {Medvedev}}]{co_hitemp}
{Li}, G., {Gordon}, I.~E., {Rothman}, L.~S., {et~al.} 2015, \apjs, 216, 15, \dodoi{10.1088/0067-0049/216/1/15}

\bibitem[{{Line} {et~al.}(2021){Line}, {Brogi}, {Bean}, {Gandhi}, {Zalesky}, {Parmentier}, {Smith}, {Mace}, {Mansfield}, {Kempton}, {Fortney}, {Shkolnik}, {Patience}, {Rauscher}, {D{\'e}sert}, \& {Wardenier}}]{Line2021}
{Line}, M.~R., {Brogi}, M., {Bean}, J.~L., {et~al.} 2021, \nat, 598, 580, \dodoi{10.1038/s41586-021-03912-6}

\bibitem[{Line {et~al.}(2021)Line, Brogi, Bean, Gandhi, Zalesky, Parmentier, Smith, Mace, Mansfield, Kempton, Fortney, Shkolnik, Patience, Rauscher, Désert, \& Wardenier}]{Line_2021}
Line, M.~R., Brogi, M., Bean, J.~L., {et~al.} 2021, Nature, 598, 580–584, \dodoi{10.1038/s41586-021-03912-6}

\bibitem[{Madhusudhan(2012)}]{Madhusudhan_2012}
Madhusudhan, N. 2012, The Astrophysical Journal, 758, 36, \dodoi{10.1088/0004-637x/758/1/36}

\bibitem[{{Molli{\`e}re} {et~al.}(2022){Molli{\`e}re}, {Molyarova}, {Bitsch}, {Henning}, {Schneider}, {Kreidberg}, {Eistrup}, {Burn}, {Nasedkin}, {Semenov}, {Mordasini}, {Schlecker}, {Schwarz}, {Lacour}, {Nowak}, \& {Schulik}}]{Molliere2022}
{Molli{\`e}re}, P., {Molyarova}, T., {Bitsch}, B., {et~al.} 2022, \apj, 934, 74, \dodoi{10.3847/1538-4357/ac6a56}

\bibitem[{{Mordasini} {et~al.}(2016){Mordasini}, {van Boekel}, {Molli{\`e}re}, {Henning}, \& {Benneke}}]{2016ApJ...832...41M}
{Mordasini}, C., {van Boekel}, R., {Molli{\`e}re}, P., {Henning}, T., \& {Benneke}, B. 2016, \apj, 832, 41, \dodoi{10.3847/0004-637X/832/1/41}

\bibitem[{{Morello} {et~al.}(2019){Morello}, {Danielski}, {Dickens}, {Tremblin}, \& {Lagage}}]{2019Morello}
{Morello}, G., {Danielski}, C., {Dickens}, D., {Tremblin}, P., \& {Lagage}, P.~O. 2019, \aj, 157, 205, \dodoi{10.3847/1538-3881/ab14e2}

\bibitem[{{{\"O}berg} {et~al.}(2011){{\"O}berg}, {Murray-Clay}, \& {Bergin}}]{2011ApJ...743L..16O}
{{\"O}berg}, K.~I., {Murray-Clay}, R., \& {Bergin}, E.~A. 2011, \apjl, 743, L16, \dodoi{10.1088/2041-8205/743/1/L16}

\bibitem[{{Padovani} \& {Cirasuolo}(2023)}]{2023ConPh..64...47P}
{Padovani}, P., \& {Cirasuolo}, M. 2023, Contemporary Physics, 64, 47, \dodoi{10.1080/00107514.2023.2266921}

\bibitem[{{Polyansky} {et~al.}(2018){Polyansky}, {Kyuberis}, {Zobov}, {Tennyson}, {Yurchenko}, \& {Lodi}}]{2018Polyansky}
{Polyansky}, O.~L., {Kyuberis}, A.~A., {Zobov}, N.~F., {et~al.} 2018, \mnras, 480, 2597, \dodoi{10.1093/mnras/sty1877}

\bibitem[{{Rothman} {et~al.}(2010){Rothman}, {Gordon}, {Barber}, {Dothe}, {Gamache}, {Goldman}, {Perevalov}, {Tashkun}, \& {Tennyson}}]{2010HITEMP}
{Rothman}, L.~S., {Gordon}, I.~E., {Barber}, R.~J., {et~al.} 2010, \jqsrt, 111, 2139, \dodoi{10.1016/j.jqsrt.2010.05.001}

\bibitem[{{Schneider} \& {Bitsch}(2021{\natexlab{a}})}]{Schneider2021.1}
{Schneider}, A.~D., \& {Bitsch}, B. 2021{\natexlab{a}}, \aap, 654, A71, \dodoi{10.1051/0004-6361/202039640}

\bibitem[{{Schneider} \& {Bitsch}(2021{\natexlab{b}})}]{Schneider2021.2}
---. 2021{\natexlab{b}}, \aap, 654, A72, \dodoi{10.1051/0004-6361/202141096}

\bibitem[{{Sim} {et~al.}(2014){Sim}, {Le}, {Pak}, {Lee}, {Kang}, {Chun}, {Jeong}, {Yuk}, {Kim}, {Park}, {Pavel}, \& {Jaffe}}]{Sim_2014}
{Sim}, C.~K., {Le}, H. A.~N., {Pak}, S., {et~al.} 2014, Advances in Space Research, 53, 1647, \dodoi{10.1016/j.asr.2014.02.024}

\bibitem[{{Snellen} {et~al.}(2010){Snellen}, {de Kok}, {de Mooij}, \& {Albrecht}}]{2010Natur.465.1049S}
{Snellen}, I. A.~G., {de Kok}, R.~J., {de Mooij}, E. J.~W., \& {Albrecht}, S. 2010, \nat, 465, 1049, \dodoi{10.1038/nature09111}

\bibitem[{{Stevenson} {et~al.}(2017{\natexlab{a}}){Stevenson}, {Line}, {Bean}, {D{\'e}sert}, {Fortney}, {Showman}, {Kataria}, {Kreidberg}, \& {Feng}}]{2017Stevenson}
{Stevenson}, K.~B., {Line}, M.~R., {Bean}, J.~L., {et~al.} 2017{\natexlab{a}}, \aj, 153, 68, \dodoi{10.3847/1538-3881/153/2/68}

\bibitem[{{Stevenson} {et~al.}(2017{\natexlab{b}}){Stevenson}, {Line}, {Bean}, {D{\'e}sert}, {Fortney}, {Showman}, {Kataria}, {Kreidberg}, \& {Feng}}]{Stevenson2017}
---. 2017{\natexlab{b}}, \aj, 153, 68, \dodoi{10.3847/1538-3881/153/2/68}

\bibitem[{Tabernero {et~al.}(2021)Tabernero, Zapatero~Osorio, Allart, Borsa, Casasayas-Barris, Demangeon, Ehrenreich, Lillo-Box, Lovis, Pallé, Sousa, Rebolo, Santos, Pepe, Cristiani, Adibekyan, Allende~Prieto, Alibert, Barros, Bouchy, Bourrier, D’Odorico, Dumusque, Faria, Figueira, Génova~Santos, González~Hernández, Hojjatpanah, Lo~Curto, Lavie, Martins, Martins, Mehner, Micela, Molaro, Nunes, Poretti, Seidel, Sozzetti, Suárez~Mascareño, Udry, Aliverti, Affolter, Alves, Amate, Avila, Bandy, Benz, Bianco, Broeg, Cabral, Conconi, Coelho, Cumani, Deiries, Dekker, Delabre, Fragoso, Genoni, Genolet, Hughes, Knudstrup, Kerber, Landoni, Lizon, Maire, Manescau, Di~Marcantonio, Mégevand, Monteiro, Monteiro, Moschetti, Mueller, Modigliani, Oggioni, Oliveira, Pariani, Pasquini, Rasilla, Redaelli, Riva, Santana-Tschudi, Santin, Santos, Segovia, Sosnowska, Spanò, Tenegi, Iwert, Zanutta, \& Zerbi}]{Tabernero_2021}
Tabernero, H.~M., Zapatero~Osorio, M.~R., Allart, R., {et~al.} 2021, Astronomy \& Astrophysics, 646, A158, \dodoi{10.1051/0004-6361/202039511}

\bibitem[{{Tsiaras} {et~al.}(2018){Tsiaras}, {Waldmann}, {Zingales}, {Rocchetto}, {Morello}, {Damiano}, {Karpouzas}, {Tinetti}, {McKemmish}, {Tennyson}, \& {Yurchenko}}]{Tsiaras2018}
{Tsiaras}, A., {Waldmann}, I.~P., {Zingales}, T., {et~al.} 2018, \aj, 155, 156, \dodoi{10.3847/1538-3881/aaaf75}

\bibitem[{Wardenier {et~al.}(2023)Wardenier, Parmentier, Line, \& Lee}]{wardenier2023modelling}
Wardenier, J.~P., Parmentier, V., Line, M.~R., \& Lee, E. K.~H. 2023, Modelling the effect of 3D temperature and chemistry on the cross-correlation signal of transiting ultra-hot Jupiters: A study of 5 chemical species on WASP-76b.
\newblock \doarXiv{2307.04931}

\bibitem[{Wardenier {et~al.}(2024)Wardenier, Parmentier, Line, Mansfield, Tan, Tsai, Bean, Birkby, Brogi, Désert, Gandhi, Lee, Levens, Pino, \& Smith}]{wardenier2024phaseresolvingabsorptionsignatureswater}
Wardenier, J.~P., Parmentier, V., Line, M.~R., {et~al.} 2024, Phase-resolving the absorption signatures of water and carbon monoxide in the atmosphere of the ultra-hot Jupiter WASP-121b with GEMINI-S/IGRINS.
\newblock \doarXiv{2406.09641}

\bibitem[{{Weaver} {et~al.}(2020){Weaver}, {L{\'o}pez-Morales}, {Espinoza}, {Rackham}, {Osip}, {Apai}, {Jord{\'a}n}, {Bixel}, {Lewis}, {Alam}, {Kirk}, {McGruder}, {Rodler}, \& {Fienco}}]{2020Weaver}
{Weaver}, I.~C., {L{\'o}pez-Morales}, M., {Espinoza}, N., {et~al.} 2020, \aj, 159, 13, \dodoi{10.3847/1538-3881/ab55da}

\bibitem[{{Weiner Mansfield} {et~al.}(2024){Weiner Mansfield}, {Line}, {Wardenier}, {Brogi}, {Bean}, {Beltz}, {Smith}, {Zalesky}, {Batalha}, {Kempton}, {Montet}, {Owen}, {Plavchan}, \& {Rauscher}}]{WM2024}
{Weiner Mansfield}, M., {Line}, M.~R., {Wardenier}, J.~P., {et~al.} 2024, arXiv e-prints, arXiv:2405.09769, \dodoi{10.48550/arXiv.2405.09769}

\bibitem[{{Welbanks} {et~al.}(2019){Welbanks}, {Madhusudhan}, {Allard}, {Hubeny}, {Spiegelman}, \& {Leininger}}]{2019Welbanks}
{Welbanks}, L., {Madhusudhan}, N., {Allard}, N.~F., {et~al.} 2019, \apjl, 887, L20, \dodoi{10.3847/2041-8213/ab5a89}

\bibitem[{{Yang} {et~al.}(2024){Yang}, {Hammond}, {Piette}, {Blecic}, {Bell}, {Irwin}, {Parmentier}, {Tsai}, {Barstow}, {Crouzet}, {Kreidberg}, {Mendon{\c{c}}a}, {Taylor}, {Baeyens}, {Ohno}, {Teinturier}, \& {Nixon}}]{Yang2024}
{Yang}, J., {Hammond}, M., {Piette}, A. A.~A., {et~al.} 2024, \mnras, \dodoi{10.1093/mnras/stae1427}

\bibitem[{{Yurchenko} {et~al.}(2020){Yurchenko}, {Mellor}, {Freedman}, \& {Tennyson}}]{Yurchenko2020}
{Yurchenko}, S.~N., {Mellor}, T.~M., {Freedman}, R.~S., \& {Tennyson}, J. 2020, \mnras, 496, 5282, \dodoi{10.1093/mnras/staa1874}

\end{thebibliography}
\bibliographystyle{aasjournal}

\end{document}